\documentclass[12pt,thmsa]{article}

\def \eq {\begin{equation}}
\def \fim-eq {\end{equation}}
\begin{document}

\author{E. S. Guerra \\
Departamento de F\'{\i}sica \\
Universidade Federal Rural do Rio de Janeiro \\
Cx. Postal 23851, 23890-000 Serop\'edica, RJ, Brazil \\
email: emerson@ufrrj.br\\
}
\title{ TELEPORTATION OF ATOMIC STATES VIA CAVITY QUANTUM ELECTRODYNAMICS}
\maketitle

\begin{abstract}
\noindent In this article we discuss a scheme of teleportation of atomic
states. The experimental realization proposed makes use of cavity Quatum
Electrodynamics involving the interaction of Rydberg atoms with a micromaser
cavity prepared in a coherent state. We start presenting a scheme to prepare
atomic Bell states via the interaction of atoms with a cavity. In our scheme
the cavity and some atoms play the role of auxiliary systems used \ to
achieve the teleportation.

\ \newline

\end{abstract}

\section{\protect\bigskip INTRODUCTION}

Entanglement and non-locality can have revolutionizing impact on our
thinking about information processing and quantum computing \cite{Nielsen,
MathQC}. Probably one of the most notable and dramatic among various
concepts developed through the application of quantum mechanics to the
information science is teleportation, put forward by Bennett \textit{et al} 
\cite{Bennett} given rise to a new field of research. The essentials of the
teleportation scheme is that, given an unknown quantum state to the sender,
making use of quantum entanglement and non-locality, it is possible \ to
reproduce this state far apart in the quantum system of the receiver where,
in the process, both the sender and the receiver follow a certain
prescription and communicate with each other through a classical channel. In
the end of the process the receiver has a quantum state similar to the
quantum state of the sender and the quantum state of the sender is destroyed
since, according to the no-cloning theorem \cite{Nocloning, Nielsen} it is
not possible to clone a quantum state. It is interesting to note that
quantum teleportation does not allow one to transmit information faster than
light in accordance with the theory of relativity because to complete the
teleportation process the sender must communicate the result of some sort of
measurement she performs to the receiver through a classical channel.
Quantum teleportation is an experimental reality and it holds tremendous
potential for applications in the fields of quantum communication and
computing \cite{MathQC, Nielsen}. For instance, it can be used \ to build
quantum gates which are resistant to noise and is intimately connected with
quantum error-correcting codes \cite{Nielsen}. The most significant
difficulty for quantum teleportation to become an useful tool in quantum
communication and computing arises in maintaining the entanglement for the
time required to transfer the classical message, that is, how to avoid
decoherence effects \cite{Orszag, Zurek}. For several proposals of
realization schemes of teleportation see \cite{MathQC}. A scheme of
teleportation of atomic states, using cavity QED, has been proposed in Ref.%
\cite{david}.

In this article we present a scheme of teleportation close to the original
scheme presented by Bennett \textit{et at } \cite{Bennett}. We will assume
that Alice and Bob meet and then create a Bell atomic state \ involving
atoms $A2$ and $\ A4$. Then Alice and Bob separate. Alice takes with her
half of the Bell pair, that is, atom $A2$ and Bob keeps with him the other
half of the Bell pair, that is, atom $A4$. Later on Alice is going to be
able to teleport to Bob's atom $A4$ an unknown state of an atom $A1$ making
use of her half of the Bell pair, that is, atom $A2$. As it will be clear,
teleportation is possible due to a fascinating and at the same time
intriguing feature of quantum mechanics: entanglement and its consequence,
non-locality. \ 

In the discussion which follows we are going to consider Rydberg atoms of
relatively long radiative lifetimes \cite{Rydat}. We also assume a perfect \
microwave cavity and we neglect effects due to decoherence. Concerning this
point, it is worth to mention that nowadays it is possible to build up
niobium superconducting cavities with high quality factors $Q$. It is
possible to construct cavities with quality factors $Q\sim 10^{8}$ \cite%
{haroche}. Even cavities with quality factors as high as $Q\sim 10^{12}$
have been reported \cite{walther}, which, for frequencies $\nu \sim 50$ GHz
gives us a cavity field lifetime of the order of a few seconds.

\section{BELL STATES}

First let us present a scheme to prepare Bell states and how to detect them.
We start assuming that we have a cavity $C$ prepared in coherent state $%
|-\alpha \rangle $. Consider a three-level cascade atom \ $Ak$ with $\mid
e_{k}\rangle ,\mid f_{k}\rangle $ and $\mid g_{k}\rangle $ being the upper,
intermediate and lower atomic states (see Fig. 1). We assume that the
transition $\mid f_{k}\rangle \rightleftharpoons \mid e_{k}\rangle $ is far
enough from resonance with the cavity central frequency such that only
virtual transitions occur between these states (only these states interact
with field in cavity $C$). In addition we assume that the transition $\mid
e_{k}\rangle \rightleftharpoons \mid g_{k}\rangle $ is highly detuned from
the cavity frequency so that there will be no coupling with the cavity
field. Here we are going to consider the effect of the atom-field
interaction taking into account only levels $\mid f_{k}\rangle $ and $\mid
g_{k}\rangle .$ We do not consider level $\mid e_{k}\rangle $ since it will
not play any role in our scheme. Therefore, we have effectively a two-level
system involving states $\mid f_{k}\rangle $ and $|g_{k}\rangle $.
Considering levels $\mid f_{k}\rangle $ and $\mid g_{k}\rangle ,$ we can
write an effective time evolution operator \cite{Gerry} 
\begin{equation}
U_{k}(t)=e^{i\varphi a^{\dagger }a}\mid f_{k}\rangle \langle f_{k}\mid
+|g_{k}\rangle \langle g_{k}\mid ,  \label{UTel}
\end{equation}%
where the second term above was put by hand just in order to take into
account the effect of level $\mid g_{k}\rangle $ and where $\varphi
=g^{2}\tau /$ $\Delta $, \ $g$ is the coupling constant, $\Delta =\omega
_{e}-\omega _{f}-\omega $ is the detuning \ where \ $\omega _{e}$ and $%
\omega _{f}$ \ are the frequencies of the upper and intermediate levels
respectively and $\omega $ is the cavity field frequency and $\tau $ is the
atom-field interaction time. Let us take $\varphi =\pi $. We assume that we
have a two-level atom $A1$ initially in the state $\mid g_{1}\rangle $, \
which is prepared in a coherent superposition according to the rotation
matrix%
\begin{equation}
R_{1}=\frac{1}{\sqrt{2}}\left[ 
\begin{array}{cc}
1 & 1 \\ 
-1 & 1%
\end{array}%
\right] ,
\end{equation}%
and we have 
\begin{equation}
\mid \psi \rangle _{A1}=\frac{1}{\sqrt{2}}(\mid f_{1}\rangle +\mid
g_{1}\rangle ).
\end{equation}%
Now, let us assume that we have a cavity $C$ prepared in coherent state $%
|-\alpha \rangle $. A coherent state $|\beta \rangle $ is obtained applying
the displacement operator $D(\beta )=e^{(\beta a^{\dag }-\beta ^{\ast }a)}$
to the vacuum, that is $|\beta \rangle =D(\beta )|0\rangle ,$ and is given
by 
\begin{equation}
|\beta \rangle =e^{-\frac{1}{2}|\beta |^{2}}{\sum\limits_{n=0}^{\infty }}%
\frac{(\beta )^{n}}{\sqrt{n!}}|n\rangle
\end{equation}%
\cite{Louisell, Orszag}. Experimentally, it is obtained with a classical
oscillating current in an antenna coupled to the cavity. Then, the system $%
A1-C$ evolves to%
\begin{equation}
\mid \psi \rangle _{A1-C}=\frac{1}{\sqrt{2}}(\mid f_{1}\rangle |\alpha
\rangle +\mid g_{1}\rangle |-\alpha \rangle ),
\end{equation}%
where we have used $e^{za^{\dagger }a}|\alpha \rangle =|e^{z}\alpha \rangle $
\cite{Louisell}$.$ Now, if atom $A1$ enters a second Ramsey cavity $R2$
where the atomic states are rotated according to the rotation matrix%
\begin{equation}
R_{2}=\frac{1}{\sqrt{2}}\left[ 
\begin{array}{cc}
1 & 1 \\ 
-1 & 1%
\end{array}%
\right] ,
\end{equation}%
we have 
\begin{eqnarray}
&\mid &f_{1}\rangle \rightarrow \frac{1}{\sqrt{2}}(\mid f_{1}\rangle -\mid
g_{1}\rangle ),  \nonumber \\
&\mid &g_{1}\rangle \rightarrow \frac{1}{\sqrt{2}}(\mid f_{1}\rangle +\mid
g_{1}\rangle ),
\end{eqnarray}%
and, therefore,%
\begin{equation}
\mid \psi \rangle _{A1-C}=\frac{1}{2}\{\mid f_{1}\rangle (|\alpha \rangle
+|-\alpha \rangle )-\mid g_{1}\rangle (|\alpha \rangle -|-\alpha \rangle )\}.
\end{equation}%
It is worth to mention at this point that if we define the non-normalized
even and odd coherent states 
\begin{eqnarray}
|+\rangle &=&|\alpha \rangle +|-\alpha \rangle ,  \nonumber \\
|-\rangle &=&|\alpha \rangle -|-\alpha \rangle ,  \label{EOCS1}
\end{eqnarray}%
with $N^{\pm }=\langle \pm \mid \pm \rangle =2\left( 1\pm e^{-2\mid \alpha
\mid ^{2}}\right) $ \ and $\langle +\mid -\rangle =0$ \cite{EvenOddCS}, we
have already a Bell state involving the atomic states of $A1$ and the cavity
field state, that is we have%
\begin{equation}
\mid \psi \rangle _{A1-C}=\frac{1}{2}(\mid f_{1}\rangle |+\rangle -\mid
g_{1}\rangle |-\rangle ).
\end{equation}

Now, let us prepare a two-level atom $A2$ in the Ramsey cavity $R3$. If atom 
$A2$ is initially in the state $\mid g_{2}\rangle $, according to the
rotation matrix%
\begin{equation}
R_{3}=\frac{1}{\sqrt{2}}\left[ 
\begin{array}{cc}
1 & 1 \\ 
-1 & 1%
\end{array}%
\right] ,
\end{equation}%
we have%
\begin{equation}
\mid \psi \rangle _{A2}=\frac{1}{\sqrt{2}}(\mid f_{2}\rangle +\mid
g_{2}\rangle ),
\end{equation}%
and let us send this atom through cavity $C$, assuming that \ for atom $A2$,
as above for atom $A1$, the transition $\mid f_{2}\rangle \rightleftharpoons
\mid e_{2}\rangle $ is far of resonance with the cavity central frequency.
Taking into account (\ref{UTel}) with $\varphi =\pi $, after the atom has
passed through the cavity we get%
\begin{eqnarray}
&\mid &\psi \rangle _{A1-A2-C}=\frac{1}{2\sqrt{2}}\{\mid f_{1}\rangle (\mid
f_{2}\rangle +\mid g_{2}\rangle )(|\alpha \rangle +|-\alpha \rangle )+ 
\nonumber \\
&\mid &g_{1}\rangle (\mid f_{2}\rangle -\mid g_{2}\rangle )(|\alpha \rangle
-|-\alpha \rangle )\}.
\end{eqnarray}%
Then, atom $A2$ enters a Ramsey cavity $R4$ where the atomic states are
rotated according to the rotation matrix%
\begin{equation}
R_{4}=\frac{1}{\sqrt{2}}\left[ 
\begin{array}{cc}
1 & 1 \\ 
-1 & 1%
\end{array}%
\right] ,
\end{equation}%
that is,%
\begin{eqnarray}
\frac{1}{\sqrt{2}}( &\mid &f_{2}\rangle +\mid g_{2}\rangle )\rightarrow \mid
f_{2}\rangle ,  \nonumber \\
\frac{1}{\sqrt{2}}( &\mid &f_{2}\rangle -\mid g_{2}\rangle )\rightarrow
-\mid g_{2}\rangle ,
\end{eqnarray}%
and we get%
\begin{equation}
\mid \psi \rangle _{A1-A2-C}=\frac{1}{2}\{\mid f_{1}\rangle \mid
f_{2}\rangle (|\alpha \rangle +|-\alpha \rangle )-\mid g_{1}\rangle \mid
g_{2}\rangle )(|\alpha \rangle -|-\alpha \rangle )\}.  \label{A1A2C}
\end{equation}%
Now, we inject $|-\alpha \rangle $ in cavity $C$ which mathematically is
represented by the operation $D(\beta )|\alpha \rangle =|\alpha +\beta
\rangle $ \cite{Louisell} and this gives us%
\begin{equation}
\mid \psi \rangle _{A1-A2-C}=\frac{1}{2}\{\mid f_{1}\rangle \mid
f_{2}\rangle (|0\rangle +|-2\alpha \rangle )-\mid g_{1}\rangle \mid
g_{2}\rangle )(|0\rangle -|-2\alpha \rangle )\}.
\end{equation}

In order to disentangle the atomic states of the cavity field state we now
send a two-level atom $A3,$ resonant with the cavity, with $|f_{3}\rangle $
and $|e_{3}\rangle $ being the lower and upper levels respectively, through $%
C$. If $A3$ is sent in the lower state $|f_{3}\rangle $, under the
Jaynes-Cummings dynamics \cite{Orszag} we know that the state $|f_{3}\rangle
|0\rangle $ does not evolve, however, the state $|f_{3}\rangle |-2\alpha
\rangle $ evolves to $|e_{3}\rangle |\chi _{e}\rangle +|f_{3}\rangle |\chi
_{f}\rangle $, where $|\chi _{f}\rangle =\sum\limits_{n}C_{n}\cos (gt\sqrt{n}%
)|n\rangle $ and $|\chi _{e}\rangle =-i\sum\limits_{n}C_{n+1}\sin (gt\sqrt{%
n+1})|n\rangle $ and $C_{n}=e^{-\frac{1}{2}|2\alpha |^{2}}(-2\alpha )^{n}/%
\sqrt{n!}$. Then we get%
\begin{eqnarray}
&\mid &\psi \rangle _{A1-A2-A3-C}=\frac{1}{2}\{\mid f_{1}\rangle \mid
f_{2}\rangle (|f_{3}\rangle |0\rangle +|e_{3}\rangle |\chi _{e}\rangle
+|f_{3}\rangle |\chi _{f}\rangle )-  \nonumber \\
&\mid &g_{1}\rangle \mid g_{2}\rangle (|f_{3}\rangle |0\rangle
-|e_{3}\rangle |\chi _{e}\rangle -|f_{3}\rangle |\chi _{f}\rangle )\},
\end{eqnarray}%
and if we detect atom $A3$ in state $|e_{3}\rangle $ finally we get the Bell
state%
\begin{equation}
\mid \Phi ^{+}\rangle _{A1-A2}=\frac{1}{\sqrt{2}}(\mid f_{1}\rangle \mid
f_{2}\rangle +\mid g_{1}\rangle \mid g_{2}\rangle ),  \label{EPRPSI+}
\end{equation}%
which is an entangled state of atoms $A1$ and $A2$, which in principle may
be far apart from each other.

In the above disentanglement process we can choose a coherent field with a
photon-number distribution with a sharp peak at average photon number $%
\langle n\rangle =|\alpha |^{2}$ so that, to a good approximation, $|\chi
_{f}\rangle \cong C_{\overline{n}}\cos (\sqrt{\overline{n}}g\tau )|\overline{%
n}\rangle $ and $|\chi _{e}\rangle \cong C_{\overline{n}}\sin (\sqrt{%
\overline{n}}g\tau )|\overline{n}\rangle $, where $\overline{n}$ is the
integer nearest $\langle n\rangle $, and we could choose, for instance $\ 
\sqrt{\overline{n}}g\tau =\pi /2$, so that we would have $|\chi _{e}\rangle
\cong C_{\overline{n}}|\overline{n}\rangle $ and $|\chi _{f}\rangle \cong 0$%
. In this case atom $A3$ \ would be detected in state $|e_{3}\rangle $ with
almost $100\%$ of probability. Therefore, proceeding this way, we can
guarantee that the atomic and field states will be disentangled successfully
as we would like.

Notice that starting from (\ref{A1A2C}) if we had injected $|\alpha \rangle $
in the cavity and detected $|e_{3}\rangle $ we would get the Bell state%
\begin{equation}
\mid \Phi ^{-}\rangle _{A1-A2}=\frac{1}{\sqrt{2}}(\mid f_{1}\rangle \mid
f_{2}\rangle -\mid g_{1}\rangle \mid g_{2}\rangle ).  \label{EPRPSI-}
\end{equation}%
\ 

Now, if we apply an extra rotation on the states of atom $A2$ in (\ref%
{EPRPSI+}) in a Ramsey cavity $R5,$ according to the rotation matrix 
\begin{equation}
R_{5}=\left[ 
\begin{array}{cc}
0 & -1 \\ 
1 & 0%
\end{array}%
\right] ,
\end{equation}%
that is,%
\begin{equation}
R_{5}=\mid f_{2}\rangle \langle g_{2}|-\mid g_{2}\rangle \langle f_{2}|,
\label{EPRR5}
\end{equation}%
we get%
\begin{equation}
\mid \Psi ^{-}\rangle _{A1-A2}=\frac{1}{\sqrt{2}}(\mid f_{1}\rangle \mid
g_{2}\rangle -\mid g_{1}\rangle \mid f_{2}\rangle ),  \label{EPRPHI-}
\end{equation}%
and applying (\ref{EPRR5}) on (\ref{EPRPSI-}) we get%
\begin{equation}
\mid \Psi ^{+}\rangle _{A1-A2}=\frac{1}{\sqrt{2}}(\mid f_{1}\rangle \mid
g_{2}\rangle +\mid g_{1}\rangle \mid f_{2}\rangle ).  \label{EPRPHI+}
\end{equation}%
The states (\ref{EPRPSI+}), (\ref{EPRPSI-}), (\ref{EPRPHI-}) and (\ref%
{EPRPHI+}) form a Bell basis \cite{BELLbasis, Nielsen} which are a complete
orthonormal basis for atoms $A1$ and $A2$.

These states show that quantum entanglement implies non-locality. The
manifestation of non-locality shows up when we perform a measurement on one
of the atoms. For instance, from (\ref{EPRPSI+}) it is clear that if we
detect atom $A1$ in state $\mid f_{1}\rangle $ then atom $A2$ collapses
instantaneously to the state $\mid f_{2}\rangle $ and if we detect atom $A1$
in state $\mid g_{1}\rangle $ then atom $A2$ collapses instantaneously to
the state $\mid g_{2}\rangle $, no matter how distant they are from each
other. The same applies to the other states (\ref{EPRPSI-}), (\ref{EPRPHI-})
and (\ref{EPRPHI+}). The Bell basis play a central role in the original
teleportation scheme proposed in \cite{Bennett}.

Now let us see how we can perform measurements in order to distinguish the
four Bell states (\ref{EPRPSI+}), (\ref{EPRPSI-}), (\ref{EPRPHI-}) and (\ref%
{EPRPHI+}). First notice that, defining 
\begin{equation}
\Sigma _{x}=\sigma _{x}^{1}\sigma _{x}^{2},
\end{equation}%
where%
\begin{equation}
\sigma _{x}^{k}=\mid f_{k}\rangle \langle g_{k}\mid +\mid g_{k}\rangle
\langle f_{k}\mid ,
\end{equation}%
we have%
\begin{eqnarray}
\Sigma _{x} &\mid &\Phi ^{\pm }\rangle _{A1-A2}=\pm \mid \Phi ^{\pm }\rangle
_{A1-A2},  \nonumber \\
\Sigma _{x} &\mid &\Psi ^{\pm }\rangle _{A1-A2}=\pm \mid \Psi ^{\pm }\rangle
_{A1-A2}.  \label{AVSIGMAx}
\end{eqnarray}%
Therefore, we can distinguish between $(\mid \Phi ^{+}\rangle _{A1-A2},\mid
\Psi ^{+}\rangle _{A1-A2})$ and $(\mid \Phi ^{-}\rangle _{A1-A2},\mid \Psi
^{-}\rangle _{A1-A2})$ performing measurements of $\Sigma _{x}=\sigma
_{x}^{1}\sigma _{x}^{2}$. In order to do so we proceed as follows. We make
use of

\begin{equation}
K_{k}=\frac{1}{\sqrt{2}}\left[ 
\begin{array}{cc}
1 & -1 \\ 
1 & 1%
\end{array}%
\right] ,
\end{equation}%
or%
\begin{equation}
K_{k}=\frac{1}{\sqrt{2}}(\mid f_{k}\rangle \langle f_{k}\mid -\mid
f_{k}\rangle \langle g_{k}\mid +\mid g_{k}\rangle \langle f_{k}\mid +\mid
g_{k}\rangle \langle g_{k}\mid ),  \label{KkEPR}
\end{equation}%
to gradually unravel the Bell states.

The eigenvectors of the operators $\sigma _{x}^{k}$ are%
\begin{equation}
|\psi _{x}^{k},\pm \rangle =\frac{1}{\sqrt{2}}(\mid f_{k}\rangle \pm \mid
g_{k}\rangle ),  \label{PSIxEPR}
\end{equation}%
and we can rewrite the Bell states as 
\begin{eqnarray}
&\mid &\Phi ^{\pm }\rangle _{A1-A2}=\frac{1}{2}[|\psi _{x}^{1},+\rangle
(\mid f_{2}\rangle \pm \mid g_{2}\rangle )+|\psi _{x}^{1},-\rangle (\mid
f_{2}\rangle \mp \mid g_{2}\rangle )],  \nonumber \\
&\mid &\Psi ^{\pm }\rangle _{A1-A2}=\frac{1}{2}[|\psi _{x}^{1},+\rangle
(\mid g_{2}\rangle \pm \mid f_{2}\rangle )+|\psi _{x}^{1},-\rangle (\mid
g_{2}\rangle \mp \mid f_{2}\rangle )].  \label{EPRPSIx}
\end{eqnarray}

Let us take for instance (\ref{EPRPSI+}) 
\begin{equation}
\mid \Phi ^{+}\rangle _{A1-A2}=\frac{1}{\sqrt{2}}(\mid f_{1}\rangle \mid
f_{2}\rangle +\mid g_{1}\rangle \mid g_{2}\rangle ).
\end{equation}%
Applying $K_{1}$ to this state we have%
\begin{equation}
K_{1}\mid \Phi ^{+}\rangle _{A1-A2}=\frac{1}{2}\{|f_{1}\rangle (\mid
f_{2}\rangle -\mid g_{2}\rangle )+|g_{1}\rangle (\mid f_{2}\rangle +\mid
g_{2}\rangle )\}.  \label{EPRK1PSI12}
\end{equation}%
Now, we compare (\ref{EPRK1PSI12}) and (\ref{EPRPSIx}). We see that the
rotation by $K_{1}$ followed by the detection of $|g_{1}\rangle $
corresponds to the detection of the the state $|\psi _{x}^{1},+\rangle $
whose eigenvalue of $\sigma _{x}^{1}$ is $+1$. After we detect $%
|g_{1}\rangle $, we get%
\begin{equation}
\mid \psi \rangle _{A2}=\frac{1}{\sqrt{2}}(\mid f_{2}\rangle +\mid
g_{2}\rangle ),
\end{equation}%
that is, we have got 
\begin{equation}
\mid \psi \rangle _{A2}=|\psi _{x}^{2},+\rangle .  \label{EPRPSI2x}
\end{equation}%
If we apply (\ref{KkEPR}) for $k=2$ to the state (\ref{EPRPSI2x}) we get%
\begin{equation}
K_{2}\mid \psi \rangle _{A2}=|g_{2}\rangle .
\end{equation}%
We see that the rotation by $K_{2}$ followed by the detection of $%
|g_{2}\rangle $ corresponds to the detection of the the state $|\psi
_{x}^{2},+\rangle $ whose eigenvalue of $\sigma _{x}^{2}$ is $+1$. The same
applies to (\ref{EPRPHI+})

We can repeat the above procedure and see that we have only 2 possibilities
which are presented schematically below, where on the left, we present the
possible sequences of atomic state rotations through $K_{k}$ and detections
of $\mid f_{k}\rangle $ or $\mid g_{k}\rangle $ and on the right, we present
the sequences of the corresponding states $|\psi _{x}^{k},\pm \rangle $
where $k=1$ and $2$ which corresponds to the measurement of the eigenvalue
of the operator $\Sigma _{x}$, $+1,$ given by (\ref{AVSIGMAx}), which
corresponds to the detection of (\ref{EPRPSI+}) or (\ref{EPRPHI+}) 
\begin{eqnarray}
(K_{1}, &\mid &g_{1}\rangle )(K_{2},\mid g_{2}\rangle )\longleftrightarrow
|\psi _{x}^{1},+\rangle |\psi _{x}^{2},+\rangle ,  \nonumber \\
(K_{1}, &\mid &f_{1}\rangle )(K_{2},\mid f_{2}\rangle )\longleftrightarrow
|\psi _{x}^{1},-\rangle |\psi _{x}^{2},-\rangle .  \label{DEPRSIGMAx+}
\end{eqnarray}

Considering (\ref{EPRPSI-}) and (\ref{EPRPHI-}) we have 
\begin{eqnarray}
(K_{1}, &\mid &g_{1}\rangle )(K_{2},\mid f_{2}\rangle )\longleftrightarrow
|\psi _{x}^{1},+\rangle |\psi _{x}^{2},-\rangle ,  \nonumber \\
(K_{1}, &\mid &f_{1}\rangle )(K_{2},\mid g_{2}\rangle )\longleftrightarrow
|\psi _{x}^{1},-\rangle |\psi _{x}^{2},+\rangle ,  \label{DEPRSIGMAx-}
\end{eqnarray}%
which corresponds to the measurement of the eigenvalue of the operator $%
\Sigma _{x}$, $-1,$ given by (\ref{AVSIGMAx}).

Now, let us see how we can make distinction between (\ref{EPRPSI+}), (\ref%
{EPRPSI-}), (\ref{EPRPHI+}) and (\ref{EPRPHI-}). For this purpose we are
going to consider (\ref{UTel}) for $\varphi =\pi $ and a cavity $C$ prepared
in the state $\mid -\alpha \rangle .$ Let us first apply $K1$ to (\ref%
{EPRPSI+}), (\ref{EPRPSI-}), (\ref{EPRPHI+}) and (\ref{EPRPHI-}), that is 
\begin{eqnarray}
K_{1} &\mid &\Phi ^{+}\rangle _{A1-A2}=\mid \Phi ^{+}\rangle _{A1-A2-K1}=%
\frac{1}{2}\{|f_{1}\rangle (\mid f_{2}\rangle -\mid g_{2}\rangle
)+|g_{1}\rangle (\mid f_{2}\rangle +\mid g_{2}\rangle )\},  \nonumber \\
K_{1} &\mid &\Phi ^{-}\rangle _{A1-A2}=\mid \Phi ^{-}\rangle _{A1-A2-K1}=%
\frac{1}{2}\{|f_{1}\rangle (\mid f_{2}\rangle +\mid g_{2}\rangle
)+|g_{1}\rangle (\mid f_{2}\rangle -\mid g_{2}\rangle )\},  \nonumber \\
K_{1} &\mid &\Psi ^{+}\rangle _{A1-A2}=\mid \Psi ^{+}\rangle _{A1-A2-K1}=%
\frac{1}{2}\{-|f_{1}\rangle (\mid f_{2}\rangle -\mid g_{2}\rangle
)+|g_{1}\rangle (\mid f_{2}\rangle +\mid g_{2}\rangle )\},  \nonumber \\
K_{1} &\mid &\Psi ^{-}\rangle _{A1-A2}=\mid \Psi ^{-}\rangle _{A1-A2-K1}=%
\frac{1}{2}\{|f_{1}\rangle (\mid f_{2}\rangle +\mid g_{2}\rangle
)-|g_{1}\rangle (\mid f_{2}\rangle -\mid g_{2}\rangle )\}  \nonumber \\
&&  \label{K1PHIPSI}
\end{eqnarray}%
Then, we pass atom $A2$ through $C$ and, taking into account (\ref{UTel}),
we get%
\begin{eqnarray}
&\mid &\Phi ^{+}\rangle _{A1-A2-K1-C}=\frac{1}{2}\{|f_{1}\rangle (\mid
f_{2}\rangle \mid \alpha \rangle -\mid g_{2}\rangle \mid -\alpha \rangle
)+|g_{1}\rangle (\mid f_{2}\rangle \mid \alpha \rangle +\mid g_{2}\rangle
\mid -\alpha \rangle )\},  \nonumber \\
&\mid &\Phi ^{-}\rangle _{A1-A2-K1-C}=\frac{1}{2}\{|f_{1}\rangle (\mid
f_{2}\rangle \mid \alpha \rangle +\mid g_{2}\rangle \mid -\alpha \rangle
)+|g_{1}\rangle (\mid f_{2}\rangle \mid \alpha \rangle -\mid g_{2}\rangle
\mid -\alpha \rangle )\},  \nonumber \\
&\mid &\Psi ^{+}\rangle _{A1-A2-K1-C}=\frac{1}{2}\{-|e_{1}\rangle (\mid
f_{2}\rangle \mid \alpha \rangle -\mid g_{2}\rangle \mid -\alpha \rangle
)+|g_{1}\rangle (\mid f_{2}\rangle \mid \alpha \rangle +\mid g_{2}\rangle
\mid -\alpha \rangle )\},  \nonumber \\
&\mid &\Psi ^{-}\rangle _{A1-A2-K1-C}=\frac{1}{2}\{|f_{1}\rangle (\mid
f_{2}\rangle \mid \alpha \rangle +\mid g_{2}\rangle \mid -\alpha \rangle
)-|g_{1}\rangle (\mid f_{2}\rangle \mid \alpha \rangle -\mid g_{2}\rangle
\mid -\alpha \rangle )\}.  \nonumber \\
&&
\end{eqnarray}%
Now, we apply a rotation on the states of $A2,$ that is, we apply 
\begin{equation}
R=\frac{1}{\sqrt{2}}\left[ 
\begin{array}{cc}
1 & 1 \\ 
-1 & 1%
\end{array}%
\right] 
\end{equation}%
which gives us%
\begin{eqnarray}
&\mid &f_{2}\rangle \rightarrow ,\frac{1}{\sqrt{2}}(\mid f_{2}\rangle -\mid
g_{2}\rangle ),  \nonumber \\
&\mid &g_{2}\rangle \rightarrow \frac{1}{\sqrt{2}}(\mid f_{2}\rangle +\mid
g_{2}\rangle ),
\end{eqnarray}%
and we get 
\begin{eqnarray}
&\mid &\Phi ^{+}\rangle _{A1-A2-K1-C-R}=\frac{1}{2\sqrt{2}}\{|f_{1}\rangle
\lbrack \mid f_{2}\rangle (\mid \alpha \rangle -\mid -\alpha \rangle )-\mid
g_{2}\rangle (\mid -\alpha \rangle +\mid \alpha \rangle )]+  \nonumber \\
|g_{1}\rangle \lbrack  &\mid &f_{2}\rangle (\mid \alpha \rangle +\mid
-\alpha \rangle )+\mid g_{2}\rangle (\mid -\alpha \rangle -\mid \alpha
\rangle )]\},  \label{K1R2PSI+} \\
&\mid &\Phi ^{-}\rangle _{A1-A2-K1-C-R}=\frac{1}{2\sqrt{2}}\{|f_{1}\rangle
\lbrack \mid f_{2}\rangle (\mid \alpha \rangle +\mid -\alpha \rangle )+\mid
g_{2}\rangle (\mid -\alpha \rangle -\mid \alpha \rangle )]+  \nonumber \\
|g_{1}\rangle \lbrack  &\mid &f_{2}\rangle (\mid \alpha \rangle -\mid
-\alpha \rangle )-\mid g_{2}\rangle (\mid \alpha \rangle +\mid -\alpha
\rangle )]\}  \label{K1R2PSI-} \\
&\mid &\Psi ^{+}\rangle _{A1-A2-K1-C-R}=\frac{1}{2\sqrt{2}}\{-|f_{1}\rangle
\lbrack \mid f_{2}\rangle (\mid \alpha \rangle -\mid -\alpha \rangle )-\mid
g_{2}\rangle (\mid -\alpha \rangle +\mid \alpha \rangle )]+  \nonumber \\
|g_{1}\rangle \lbrack  &\mid &f_{2}\rangle (\mid \alpha \rangle +\mid
-\alpha \rangle )+\mid g_{2}\rangle (\mid -\alpha \rangle -\mid \alpha
\rangle )]\},  \label{K1R2PHI+} \\
&\mid &\Psi ^{-}\rangle _{A1-A2-K1-C-R}=\frac{1}{2\sqrt{2}}\{|f_{1}\rangle
\lbrack \mid f_{2}\rangle (\mid \alpha \rangle +\mid -\alpha \rangle )+\mid
g_{2}\rangle (\mid -\alpha \rangle -\mid \alpha \rangle )]+  \nonumber \\
-|g_{1}\rangle \lbrack  &\mid &f_{2}\rangle (-\mid \alpha \rangle +\mid
-\alpha \rangle )+\mid g_{2}\rangle (\mid \alpha \rangle +\mid -\alpha
\rangle )]\}  \label{K1R2PHI-}
\end{eqnarray}%
Now, for (\ref{K1R2PSI+}) we displace the cavity injecting $\mid \alpha
\rangle $ and send a two-level atom $A3$ resonant with the cavity through $C$%
. If $A3$ is sent in the lower state $|f_{3}\rangle $ and after it crosses
the cavity we detect the upper $|e_{3}\rangle $ we get%
\begin{equation}
\mid \Phi ^{+}\rangle _{A1-A2}=\frac{1}{2}\{(|f_{1}\rangle +|g_{1}\rangle
)(\mid f_{2}\rangle -\mid g_{2}\rangle )\},
\end{equation}%
and if we apply (\ref{KkEPR}) to atoms $A1$ and $A2$ we get 
\begin{equation}
\mid \Phi ^{+}\rangle _{A1-A2-K1-K2}=|g_{1}\rangle \mid f_{2}\rangle .
\end{equation}%
For (\ref{K1R2PSI-}) we displace the cavity injecting $\mid -\alpha \rangle $
and, as above, sending a two-level atom $A3$ through $C$ in the lower state $%
|f_{3}\rangle $ and after it crosses the cavity detecting the upper state $%
|e_{3}\rangle $ we get%
\begin{equation}
\mid \Phi ^{-}\rangle _{A1-A2}=\frac{1}{2}\{(|f_{1}\rangle -|g_{1}\rangle
)(\mid f_{2}\rangle +\mid g_{2}\rangle )\},
\end{equation}%
and if we apply (\ref{KkEPR}) to atoms $A1$ and $A2$ we get 
\begin{equation}
\mid \Phi ^{-}\rangle _{A1-A2-K1-K2}=|f_{1}\rangle \mid g_{2}\rangle .
\end{equation}%
For (\ref{K1R2PHI+}) we displace the cavity injecting $\mid \alpha \rangle $
and sending a two-level atom $A3$ through $C$ in the lower state $%
|f_{3}\rangle $ and after it crosses the cavity detecting the upper state $%
|e_{3}\rangle ,$ we get%
\begin{equation}
\mid \Psi ^{+}\rangle _{A1-A2}=\frac{1}{2}\{(-|f_{1}\rangle +|g_{1}\rangle
)(\mid f_{2}\rangle -\mid g_{2}\rangle )\},
\end{equation}%
and if we apply (\ref{KkEPR}) to atoms $A1$ and $A2$ we get 
\begin{equation}
\mid \Psi ^{+}\rangle _{A1-A2-K1-K2}=|f_{1}\rangle \mid f_{2}\rangle .
\end{equation}%
And finally, for (\ref{K1R2PHI-}) we displace the cavity injecting $\mid
-\alpha \rangle $ and sending a two-level atom $A3$ through $C$ in the lower
state $|f_{3}\rangle $ and after it crosses the cavity detecting the upper
state $|e_{3}\rangle ,$ we get%
\begin{equation}
\mid \Psi ^{-}\rangle _{A1-A2}=\frac{1}{2}\{(|f_{1}\rangle +|g_{1}\rangle
)(\mid f_{2}\rangle +\mid g_{2}\rangle ),
\end{equation}%
and if we apply (\ref{KkEPR}) to atoms $A1$ and $A2$ we get finally%
\begin{equation}
\mid \Psi ^{-}\rangle _{A1-A2-K1-K2}=|g_{1}\rangle \mid g_{2}\rangle .
\end{equation}

As we see, the discrimination between (\ref{EPRPSI+}), (\ref{EPRPSI-}), (\ref%
{EPRPHI+}) and (\ref{EPRPHI-}), is made detecting, after the process
described above,( $|g_{1}\rangle \mid f_{2}\rangle ),(|f_{1}\rangle \mid
g_{2}\rangle ),(|f_{1}\rangle \mid f_{2}\rangle $ ) or ( $|g_{1}\rangle \mid
g_{2}\rangle ).$

\section{TELEPORTATION}

In this section we are going discuss a teleportation scheme that is closely
similar to the original scheme suggested by Bennett \textit{et al} \cite%
{Bennett}. Let us assume that Alice and Bob meet and than they build up a
Bell state involving two-level atoms $A2$ and $A4$ as described in section 2
(we use the notation $A3$ for the two-level atom used to disentangle the
atomic states from the cavity state as in the previous section). That is, as
\ in \ the previous section they make use of a cavity prepared initially in
a coherent state and send $A2$ and $A4$ through this cavity where the atoms
interact dispersively with the cavity, and following the recipe presented in
that section, they get

\begin{equation}
\mid \Phi ^{+}\rangle _{A2-A4}=\frac{1}{\sqrt{2}}(\mid f_{2}\rangle \mid
f_{4}\rangle +\mid g_{2}\rangle \mid g_{4}\rangle ),
\end{equation}%
Now, let us assume that Alice keeps with her the half of this Bell state
consisting of atom $A2$ and Bob keeps with him the other half of this Bell
state, that is, atom $A4$. Then they separate and let us assume that they
are far apart from each other. Later on, Alice decides to teleport the state
of an atom $A1$ prepared in an unknown state%
\begin{equation}
\mid \psi \rangle _{A1}=\zeta \mid f_{1}\rangle +\xi \mid g_{1}\rangle
\label{TPSIA1}
\end{equation}%
to Bob. For this purpose let us write \ the state formed by the direct
product of the Bell state and this unknown state $\mid \Psi ^{+}\rangle
_{A2-A4}\mid \psi \rangle _{A1}$, that is,%
\begin{eqnarray}
&\mid &\psi \rangle _{A1-A2-A4}=\frac{1}{\sqrt{2}}\{\zeta (\mid f_{1}\rangle
\mid f_{2}\rangle \mid f_{4}\rangle +\mid f_{1}\rangle \mid g_{2}\rangle
\mid g_{4}\rangle )+  \nonumber \\
\xi ( &\mid &g_{1}\rangle \mid f_{2}\rangle \mid f_{4}\rangle +\mid
g_{1}\rangle \mid g_{2}\rangle \mid g_{4}\rangle ).  \label{TELPSIA1A2A4}
\end{eqnarray}

First Alice prepares a cavity $C$ in a coherent state $\mid -\alpha \rangle $%
. Taking into account (\ref{UTel}) with $\varphi =\pi ,$ after atoms $A1$
and $A2$ fly through the cavity we have%
\begin{eqnarray}
&\mid &\psi \rangle _{A1-A2-A4-C}=\frac{1}{\sqrt{2}}\{\zeta (\mid
f_{1}\rangle \mid f_{2}\rangle \mid -\alpha \rangle \mid f_{4}\rangle +\mid
f_{1}\rangle \mid g_{2}\rangle \mid \alpha \rangle \mid g_{4}\rangle )+ 
\nonumber \\
\xi ( &\mid &g_{1}\rangle \mid f_{2}\rangle \mid \alpha \rangle \mid
f_{4}\rangle +\mid g_{1}\rangle \mid g_{2}\rangle \mid -\alpha \rangle \mid
g_{4}\rangle )\}.  \label{TPSIA1A2A4C}
\end{eqnarray}%
Now, we make use of the Bell basis involving atom $A1$ and $A2$ and we can
write 
\begin{eqnarray}
&\mid &f_{1}\rangle \mid f_{2}\rangle =\frac{1}{\sqrt{2}}(\mid \Phi
^{+}\rangle _{A1-A2}+\mid \Phi ^{-}\rangle _{A1-A2}),  \nonumber \\
&\mid &g_{1}\rangle \mid g_{2}\rangle =\frac{1}{\sqrt{2}}(\mid \Phi
^{+}\rangle _{A1-A2}-\mid \Phi ^{-}\rangle _{A1-A2}),  \nonumber \\
&\mid &f_{1}\rangle \mid g_{2}\rangle =\frac{1}{\sqrt{2}}(\mid \Psi
^{+}\rangle _{A1-A2}+\mid \Psi ^{-}\rangle _{A1-A2}),  \nonumber \\
&\mid &g_{1}\rangle \mid f_{2}\rangle =\frac{1}{\sqrt{2}}(\mid \Psi
^{+}\rangle _{A1-A2}-\mid \Psi ^{-}\rangle _{A1-A2}).
\end{eqnarray}%
Therefore, we can rewrite (\ref{TPSIA1A2A4C}) as%
\begin{eqnarray}
&\mid &\psi \rangle _{A1-A2-A4-C}=  \nonumber \\
\frac{1}{\sqrt{2}}\{ &\mid &\Phi ^{+}\rangle _{A1-A2}[\zeta \mid
f_{4}\rangle +\xi \mid g_{4}\rangle ]\mid -\alpha \rangle +  \nonumber \\
&\mid &\Phi ^{-}\rangle _{A1-A2}[\zeta \mid f_{4}\rangle -\xi \mid
g_{4}\rangle ]\mid -\alpha \rangle +  \nonumber \\
&\mid &\Psi ^{+}\rangle _{A1-A2}[\zeta \mid g_{4}\rangle +\xi \mid
f_{4}\rangle ]\mid \alpha \rangle +  \nonumber \\
&\mid &\Psi ^{-}\rangle _{A1-A2}[\zeta \mid g_{4}\rangle -\xi \mid
f_{4}\rangle ]\mid \alpha \rangle \}.  \label{PSIA1A2A4CEPR}
\end{eqnarray}%
Notice that atoms $A1$ and $A2$ are with Alice and she wants to teleport the
state of atom $A1$ (\ref{TPSIA1}) to Bob's atom $A4$. Inspecting (\ref%
{PSIA1A2A4CEPR}) we see that all Alice has to do is to inject in the cavity $%
\mid \alpha \rangle $ or $\mid -\alpha \rangle $ and to perform a
measurement of one of the Bell state that form the Bell basis. We have
already seen how to detect the Bell states in section 2. Then proceeding
according the prescription detailed in section 2 Alice, in the end, has four
possibilities. First let us assume that Alice injects in the cavity $\mid
-\alpha \rangle $. \ Then, she sends a two-level atom $A3$ resonant with the
cavity in the lower state $\mid f_{3}\rangle $ and after it crosses the
cavity she detects the upper state $\mid e_{3}\rangle $ and she gets%
\begin{equation}
\mid \psi \rangle _{A1-A2-A4}=\frac{1}{N}\{\mid \Phi ^{+}\rangle
_{A1-A2}[\zeta (\mid f_{4}\rangle +\xi \mid g_{4}\rangle ]+\mid \Phi
^{-}\rangle _{A1-A2}[\zeta \mid f_{4}\rangle -\xi \mid g_{4}\rangle ]\},
\end{equation}%
where $N$ is a normalization constant. Then, she has four alternatives, that
is, applying \ (\ref{KkEPR})\ \ to $A1$ and $A2$ respectively, according to (%
\ref{DEPRSIGMAx+}) and (\ref{DEPRSIGMAx-}) if she gets $(K_{1},\mid
g_{1}\rangle )(K_{2},\mid g_{2}\rangle )$ or $(K_{1},\mid f_{1}\rangle
)(K_{2},\mid f_{2}\rangle )$ this corresponds to the detection \ of $\mid
\Phi ^{+}\rangle _{A1-A2}$ and if she gets $(K_{1},\mid f_{1}\rangle
)(K_{2},\mid g_{2}\rangle )$ or $(K_{1},\mid g_{1}\rangle )(K_{2},\mid
f_{2}\rangle )$ this corresponds to the detection \ of $\mid \Phi
^{-}\rangle _{A1-A2}$. Therefore, after the detection of the states of the
atoms $A1$ and $A2$ Alice calls Bob and informs him that she has injected $%
\mid -\alpha \rangle $ \ in the cavity and the result of her atomic
detection so that Bob knows what to do to get the right state, that is, an
state similar to (\ref{TPSIA1}). If she detects $(\mid f_{1}\rangle \mid
f_{2}\rangle )$ or ($\mid g_{1}\rangle \mid g_{2}\rangle )$ Bob gets 
\begin{equation}
\mid \psi \rangle _{A4}=\zeta \mid f_{4}\rangle +\xi \mid g_{4}\rangle ,
\label{PSIright}
\end{equation}%
and he has to do nothing else. If she detects $(\mid f_{1}\rangle \mid
g_{2}\rangle )$ \ or $(\mid g_{1}\rangle \mid f_{2}\rangle )$ \ Bob gets%
\begin{equation}
\mid \psi \rangle _{A4}=\zeta \mid f_{4}\rangle -\xi \mid g_{4}\rangle ,
\end{equation}%
and he must apply a rotation in the Ramsey cavity $R4$ 
\begin{equation}
R_{4}=\left[ 
\begin{array}{cc}
1 & 0 \\ 
0 & -1%
\end{array}%
\right] ,
\end{equation}%
in order to get a state like (\ref{PSIright}).

Now let us assume that Alice injects in the cavity $\mid \alpha \rangle $. \
Then she sends a two-level atom $A3$ resonant with the cavity in the lower
state $\mid f_{3}\rangle $ and after it crosses the cavity she detects the
upper state $\mid e_{3}\rangle $ and she gets 
\begin{equation}
\mid \psi \rangle _{A1-A2-A4}=\frac{1}{N}\{\mid \Psi ^{+}\rangle
_{A1-A2}[\zeta \mid g_{4}\rangle +\xi \mid f_{4}\rangle ]+\mid \Psi
^{-}\rangle _{A1-A2}[\zeta \mid g_{4}\rangle -\xi \mid f_{4}\rangle ]\}.
\end{equation}%
Again she has four alternatives, that is, applying \ (\ref{KkEPR})\ \ to $A1$
and $A2$ respectively, according to (\ref{DEPRSIGMAx+}) \ and (\ref%
{DEPRSIGMAx-}) if she gets $(K_{1},\mid g_{1}\rangle )(K_{2},\mid
g_{2}\rangle )$ or $(K_{1},\mid f_{1}\rangle )(K_{2},\mid f_{2}\rangle )$
this corresponds to the detection \ of $\mid \Psi ^{+}\rangle _{A1-A2}$ and
if she gets $(K_{1},\mid f_{1}\rangle )(K_{2},\mid g_{2}\rangle )$ or $%
(K_{1},\mid g_{1}\rangle )(K_{2},\mid f_{2}\rangle )$ this corresponds to
the detection \ of $\mid \Psi ^{-}\rangle _{A1-A2}$. Therefore, after the
detection of the states of the atoms $A1$ and $A2$, Alice calls Bob and
informs him that she has injected $\mid \alpha \rangle $ \ in the cavity and
the result of her atomic detection to Bob. If she detects$(\mid f_{1}\rangle
\mid f_{2}\rangle $ $)$\ or $(\mid g_{1}\rangle \mid g_{2}\rangle )$ Bob gets%
\begin{equation}
\mid \psi \rangle _{A4}=\xi \mid f_{4}\rangle +\zeta \mid g_{4}\rangle ,
\end{equation}%
and he has to apply a rotation in the Ramsey cavity $R4$ 
\begin{equation}
R_{4}=\left[ 
\begin{array}{cc}
0 & 1 \\ 
1 & 0%
\end{array}%
\right] ,
\end{equation}%
to get (\ref{PSIright}). Finally, if she detects $(\mid f_{1}\rangle \mid
g_{2}\rangle )$ or $(\mid g_{1}\rangle \mid f_{2}\rangle )$ Bob gets 
\begin{equation}
\mid \psi \rangle _{A4}=-\xi \mid f_{4}\rangle +\zeta \mid g_{4}\rangle ,
\end{equation}%
and he has to apply a rotation in the Ramsey cavity $R4$ 
\begin{equation}
R_{4}=\left[ 
\begin{array}{cc}
0 & 1 \\ 
-1 & 0%
\end{array}%
\right] ,
\end{equation}%
to get (\ref{PSIright}). Notice that the original state (\ref{TPSIA1}) \ is
destroyed in the end of the teleportation process (it evolves to \ $\mid
f_{1}\rangle $ or $\mid g_{1}\rangle $) in accordance with the no-cloning
theorem \cite{Nielsen, Nocloning}.

In Fig. 2 we present the scheme of the teleportation process we have
discussed above.

Concluding, we have presented a scheme of realization of atomic state
teleportation making use of cavity QED. A nice alternative scheme also
making use of atoms interacting with electromagnetic cavities has also been
proposed in Ref. \cite{david}. In our scheme we use atoms interacting with
superconducting cavities prepared in a coherent state which are states
relatively easy to be prepared and handled. In Ref. \cite{david} it is used
atoms interacting with cavities prepared in Fock states which are state of
the electromagnetic field which are sensitive to decoherence. We think that
both this schemes could be realized experimentally in the future.

\bigskip

\bigskip \textbf{Figure Captions} \newline
\newline

\textbf{Fig. 1-} Energy states scheme of a three-level atom where $|e\rangle 
$ is the upper state with atomic frequency $\omega _{e}$, $\ |f\rangle $ is
the intermediate state with atomic frequency $\omega _{f}$, $|g\rangle $ is
the lower state with atomic frequency $\omega _{g}$ and $\omega $ is the
cavity field frequency and $\Delta =(\omega _{e}-\omega _{f})-\omega $ is
the detuning.

\bigskip

\textbf{Fig. 2- }Set-up for teleportation process. Alice and Bob meet and
generate a Bell state involving atoms $A2$ and $A4$. Alice sends atoms $A1$
and $A2$ through a cavity $C$ prepared initially in a coherent state $%
|-\alpha \rangle $. After atoms $A1$ and $A2$ have flown through $C$ Alice
must inject $|\alpha \rangle $ or $|-\alpha \rangle $ \ in the cavity, send
a two-level atom $A3$ resonant with the cavity through $C$ in the lower
state $|f_{3}\rangle $ and detect the upper state $|e_{3}\rangle .$ Then she
must perform a measurement of the remaining Bell states of the Bell basis.
For this purpose she sends atom $A1$ through the Ramsey cavity $K_{1}$ and $%
A2$ through Ramsey cavity $K_{2}$. Then, she calls Bob and informs him which
coherent field she has injected in $C1$ and the result of her atomic
detections in detectors $D1$ and $D2$. Depending on the results of the
Alice's atomic detections and which coherent state she injected in the
cavity, Bob has or not to perform an extra rotation in the Ramsey cavity $R4$
on the states of his atom $A4.$


\bigskip

\begin{thebibliography}{99}
\bibitem{Nielsen} M. A. Nielsen and I. L. Chuang, \textit{Quantum
Computation and Quantum Information}, (Cambridge Univ. Press, Cambridge,
2000).

\bibitem{MathQC} G. Chen and R. K. Brylinski Eds., \textit{Mathematics of
Quantum Computation}, (Chapman \& Hall/CRC, London, 2002).

\bibitem{Bennett} C. H. Bennett, C. Brassard, C. Cr\'{e}peau, R. Jozsa, A.
Peres and W. K. Wootters, Phys. Rev. Lett. \textbf{70},\textbf{\ }1895%
\textbf{\ }(1993).

\bibitem{Nocloning} W. K. Wootters and W. H. Zurek, Nature \textbf{299}, 802
(1982).

\bibitem{Orszag} M. Orszag, \textit{Quantum Optics}, (Springer-Verlag,
Berlin, 2000).

\bibitem{Zurek} W. H. Zurek, Phys. Today, October \textbf{44}, 36 (1991);
Phys. Rev. D \textbf{24}, 1516 (1981); Phys. Rev. D \textbf{26}, 1862 (1982).

\bibitem{david} L. Davidovich, N. Zagury, M. Brune, J. M. Raimond and S.
Haroche, Phys. Rev A \textbf{50} (1994), R895.

\bibitem{Rydat} A. A. Radzig and B. M. Smirnov, \textit{Reference Data on
Atoms, Molecules, and Ions} (Springer-Verlag, Berlin, 1985); T. F.
Gallagher, \textit{Rydberg Atoms }(Cambridge Univ. Press, Cambridge, 1984).

\bibitem{haroche} M. Brune, F. Schmidt-Kaler, A. Mauli, J. Dreyer, E.
Hagley, J. M. Raimond, and S. Haroche, Phys. Rev. Lett. \textbf{76}, 1800
(1996).

\bibitem{walther} G. Rempe, F. Schmidt-Kaler and H. Walther, Phys. Rev.
Lett. \textbf{64},\textbf{\ }2783 (1990).

\bibitem{Gerry} C. C. Gerry, Phys. Rev. A \textbf{53}, 2857 (1996); Phys.
Rev. A \textbf{54}, R2529 (1996).

\bibitem{Louisell} W. H. Louisell, \textit{Quantum Statistical Properties of
Radiation}, (Wiley,\textbf{\ }New York, 1973).

\bibitem{EvenOddCS} E. S. Guerra, B. M. Garraway and P. L. Knight, Phys.
Rev. A \textbf{55}, 3482 (1997).

\bibitem{BELLbasis} S. L. Braustein, A. Mann and M. Revzen, Phys. Rev. Lett. 
\textbf{68}, 3259 (1992).
\end{thebibliography}
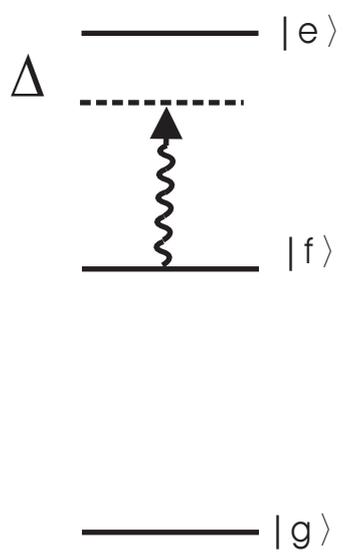
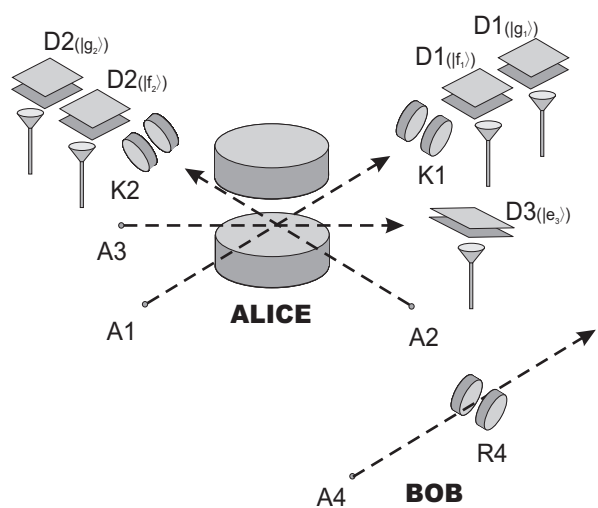
\end{document}